\documentclass{elsart}

\usepackage{amssymb}
\usepackage{amsmath}
\usepackage{graphicx}
\usepackage{cite}
\usepackage{bm}
\usepackage{epsfig}
\begin{document}
\begin{frontmatter}


\title{ Goodness of fit tests for  weighted histograms}
\author{N.D. Gagunashvili\thanksref{now}}
\thanks[now]{Present address: Max-Planck-Institut f\"{u}r Kernphysik, PO Box
103980, \\69029 Heidelberg, Germany} \ead{nikolai@unak.is}
\address {University of Akureyri, Borgir, v/Nordursl\'od, IS-600 Akureyri, Iceland }
\begin{abstract}
Weighted histogram in Monte-Carlo simulations is often used for the
estimation of a probability density function. It is obtained as a
result of random experiment with random events that have weights. In
this paper the bin contents of weighted histogram are considered as
a sum of random variables with random number of terms. Goodness of
fit tests for weighted histograms and for weighted histograms with
unknown normalization are proposed. Sizes and powers of the tests
are investigated numerically.
\end{abstract}
\end{frontmatter}
\section{Introduction}
 A histogram with $m$ bins for a given probability density function  $p(x)$ is used to estimate the
 probabilities
\begin{equation}
p_i=\int_{S_i}p(x)dx, \; i=1,\ldots ,m \label{p1}
\end{equation}
 that a random event  belongs to  bin $i$. Integration in (\ref{p1}) is done over the bin  $S_i$.

A histogram can be obtained as a result of a random experiment with probability
 density function $p(x)$.
Let us denote the number of random events belonging to the $i$th bin
of the  histogram as $n_{i}$. The total number of events in the
histogram is equal to $n=\sum_{i=1}^{m}{n_i}$.  The quantity
$\hat{p}_i= n_{i}/n$ is an estimator
 of $p_i$  with expectation value $\textrm E \,\hat{p_i}=p_i$. The distribution
  of the number of events for  bins of the  histogram is the multinomial
   distribution \cite{agresti}  and the probability of the random vector
     $(n_1,\ldots ,n_m)$  is given by
\begin{equation}
P(n_1,\ldots ,n_m)=\frac{n!}{n_1!n_2! \ldots n_m!} \; p_1^{n_1}
\ldots p_m^{n_m},\text{ } \sum_{i=1}^{m} p_i=1.
\end{equation}
The problem of goodness of fit is to test the hypothesis
\begin{equation}
H_0: p_1=p_{10},\ldots, p_{m-1}=p_{m-1,0} \text{  vs.  } H_a: p_i \neq p_{i0} \text{  for some  } i,
\end{equation}
where $p_{i0}$ are specified probabilities, and $\sum_{i=1}^{m}
p_{i0}=1$. The test is used in a data  analyses for comparison
theoretical frequencies $np_{i0}$ with the observed frequencies
$n_i$. This classical problem remains of current practical interest.
The test statistic
\begin{equation}
X^2=\sum_{i=1}^{m} \frac{(n_i-np_{i0})^2}{np_{i0}} \label{basic}
\end{equation}
was suggested by Pearson \cite{pearson}. Pearson showed that the
statistic (\ref{basic}) has  approximately a $\chi^2_{m-1}$
distribution if the hypothesis
 $H_0$ is true.
Improvements of the chi-square test were proposed in \cite{yates,
Hoel,Nass}, also known are the likelihood ratio test \cite{Dunning}
and an exact test  \cite{tate}. Review and comparison of different
multinomial goodness-of-fit tests was done in \cite {cressie}, a
detailed numerical investigation has been given in \cite{Krishna}.

Weighted histograms are often obtained as a result of Monte-Carlo
simulations. References \cite{muon,weight,astro} are examples of
research works in high energy physics, statistical mechanics and
astrophysics using such histograms. Operations with weighted
histograms are realised in contemporary systems for data analysis
HBOOK \cite{hbook}, Physics Analysis Workstation(PAW)\cite{paw} and
ROOT framework \cite{root}, developed at CERN (European Organization
for Nuclear Research, Geneva, Switzerland).

To define a weighted histogram let us write the probability $p_i$
(\ref{p1}) for a given probability density function  $p(x)$  in the
form
\begin{equation}
p_i= \int_{S_i}p(x)dx = \int_{S_i}w(x)g(x)dx, \label{weightg}
\end{equation}
where
\begin{equation}
w(x)=p(x)/g(x) \label{fweight}
\end{equation}
 is the weight function and $g(x)$ is some other probability density function. The function $g(x)$ must be larger
 than 0 for points $x$, where $p(x)\neq 0$. Weight $w(x)=0$ if $p(x)=0$ \cite{Sobol}.

The weighted histogram is obtained as a result of a random experiment with probability
density function $g(x)$ and weights of events calculated according to (\ref{fweight}).
Let us denote the total sum of weights of events  in the $i$th  bin of the weighted histogram as
\begin{equation}
W_i= \sum_{k=1}^{n_i}w_i(k), \label{ffweight}
\end{equation}
where $n_i$ is the number of events at bin $i$ and $w_i(k)$ is the
weight of $k$th event in the $i$th bin. The total number of events
in the histogram is equal to $n=\sum_{i=1}^{m}{n_i}$, where $m$ is
the number of bins. The quantity $\hat{p}_i= W_{i}/n$ is the
estimator of $p_i$ with expectation value $\textrm E \,
\hat{p_i}=p_i$.
 Notice that in the  case when $g(x)=p(x)$ the weights of  events are equal to 1 and the weighted histogram
 is the usual histogram.
 For weighted histograms again the problem of goodness of fit is to test the hypothesis
\begin{equation}
H_0: p_1=p_{10},\ldots ,p_{m-1}=p_{m-1,0} \text{  vs.  } H_a: p_i \neq p_{i0} \text{  for some  } i,
\end{equation}
where $p_{i0}$ are specified probabilities, and $\sum_{i=1}^{m} p_{i0}=1$.

In  practice the heuristic "chi-square" test statistic is used for
this purpose
\begin{equation}
X^2_h=\sum_{i=1}^{m} \frac{(W_i-np_{i0})^2}{W_{2i}},
\end{equation}
where
\begin{equation}
W_{2i}= \sum_{k=1}^{n_i}w_i(k)^2. \label{fffweight}
\end{equation}
It is expected that if hypothesis $H_0$ is true then statistic
$X^2_h$ has $\chi^2_{m-1}$ distribution. The recommended minimal
number of events in a bin is equal to 25 for application this test
\cite{hbook,paw,root}.

 The next section of this paper proposes a generalization of the chi-square
  test for weighted histograms, a goodness of fit test for weighted
histograms with unknown normalization is proposed in section 3. To
evaluate the tests, in section  4   the sizes and powers of the
tests are calculated for numerical examples with different numbers
of events, bins and weight functions.
 The size of the test is compared with the  calculated size of the heuristic
  chi-square test.  The comparison demonstrates the   superiority of the
   proposed  generalization of chi-square test over the heuristic chi-square test.

\section{The test}

The total sum of weights of events in $i$th  bin  $W_{i}$,
$i=1,\ldots,m$ can be  considered as a sum of random variables
 \begin{equation}
W_i= \sum_{k=1}^{n_i}w_i(k), \label{ffffweight}
\end{equation}
where also the number of events  $n_i$ is a random value and the
weights $w_i(k),k=1,...,n_i$ are independent
 random variables with the same probability distribution function.
 The distribution of the number of events for bins of the histogram  is
  the multinomial distribution  and the probability of the random vector
  $(n_1,\ldots ,n_m)$ is
\begin{equation}
P(n_1,\ldots ,n_m)=\frac{n!}{n_1!n_2! \ldots n_m!} \; g_1^{n_1}
\ldots g_m^{n_m},\text{ } \sum_{i=1}^{m} g_i=1,
\end{equation}
where
\begin{equation}
g_i=\int_{S_i}g(x)dx, \; i=1,\ldots ,m \label{p}
\end{equation}
  is the probability that a random event  belongs to the  bin $i$. Integration
   in (\ref{p}) is done over the bin  $S_i$.

 Let us denote the expectation values of the weights of events from the $i$th bin
  as $\textrm E \,w_{i}= \mu_i$ and the
  variances as $\textrm {Var} \, w_{i}= \sigma_i^2$. The expectation value
   of the total sum of weights $W_{i}, i=1,\ldots,m$ is \cite{gnedenko}:
\begin{equation}
\textrm E \,W_i= \textrm E \, \sum_{k=1}^{n_i}w_{i}(k)=  \textrm E\,
w_{i} \textrm E \,n_i =  n\mu_ig_i.\label{eew}
\end{equation}
The diagonal elements $\gamma_{ii}$ of the  covariance matrix of the
vector $(W_1,\ldots ,W_m)$  are equal to \cite{gnedenko}
\begin{equation}
\gamma_{ii}=\sigma_i^2 g_in+\mu_i^2 g_i(1-g_i)n=
n\alpha_{2i}g_i-n\mu_i^2g_i^2, \label{gii}
\end{equation}
where $\alpha_{2i}= \textrm E \, w_{i}^2$. The non-diagonal elements
$\gamma_{ij},\,i\neq j$  are equal to:
\begin{equation}
\begin{split}
\gamma_{ij}=\sum_{k=0}^n\sum_{l=0}^{n}\textrm E
\,[\sum_{u=1}^k\sum_{v=1}^l w_i(u)w_j(v)]h(k,l)
-\textrm E \,W_i\textrm E \,W_j\\
=\sum_{k=0}^n\sum_{l=0}^{n}\textrm E \,(w_{i}w_{j})h(k,l)kl-\mu_ing_i\mu_jng_j\quad\quad\quad\quad\,\\
=\mu_i\mu_j(-g_ig_jn+g_ig_jn^2 )-
\mu_ing_i\mu_jng_j\quad\quad\quad\quad\;\:\,\,\\=-n\mu_i\mu_j
g_ig_j,\quad\quad\quad\quad\quad\quad\quad\quad\quad\quad\quad\quad\quad\quad\quad\,
 \label{gij}
\end{split}
\end{equation}
where $h(k,l)$ is the probability that $k$ events belong to bin $i$
and $l$ events to bin $j$.

 If hypothesis $H_0$ is true then
\begin{equation}
\textrm E \,W_i=n\mu_ig_i=np_{i0}, \; i=1,\ldots,m \label{ew}
\end{equation}
and
\begin{equation}
g_i=p_{i0}/\mu_i, \; i=1,\ldots,m. \label{pi}
\end{equation}
We can substitute $g_i$ to  (\ref{gii}) which gives
\begin{equation}
\gamma_{ii}= n(\frac {p_{i0}}{r_i} -p_{i0}^2),\label{ga11}
\end{equation}
where $r_i=\mu_i/ \alpha_{2i}$.  Substituting $g_i$ into (\ref{gij}) gives
\begin{equation}
\gamma_{ij}=-np_{i0}p_{j0}.
\end{equation}
Notice that for usual histograms the ratio of moments $r_i$ is equal
to 1 and the covariance matrix coincides with the covariance matrix
of the multinomial distribution.

Let us now introduce the multivariate $T^2$ Hotelling statistic
\begin{equation}
(\textbf{W}-n\textbf{p}_0)'\Gamma_k^{-1}(\textbf{W}-n\textbf{p}_0),
\end{equation}
where \\ $\textbf{W}=(W_1,\ldots,W_{k-1},W_{k+1},\ldots,W_{m})'$,
$\textbf{p}_0=(p_{10},\ldots,p_{k-1,0},p_{k+1,0},\ldots,p_{m0})'$
and $\Gamma_k=(\gamma_{ij})_{(m-1)\times(m-1)}$ is the covariance
matrix for a histogram without bin $k$. The matrix $\Gamma_k$ has
the form
\begin{equation}
\Gamma_k=\textrm{diag}\,(n\frac{p_{10}}{r_1}, \ldots
,n\frac{p_{k-1,0}}{r_{k-1}},n\frac{p_{k+1,0}}{r_{k+1}}, \ldots ,
n\frac{p_{m 0}}{r_m})-n \textbf{p}_0\textbf{p}_0',
\end{equation}
and the Woodbury theorem \cite{woodbury} can be applied to find
$\Gamma_k^{-1}$. After that the Hotelling statistic can be written
as
\begin{equation}
 X^2_k= \sum_{i \neq k} r_i \frac{(W_i-np_{i0})^2}{np_{i0}}+\frac{(\sum_{i \neq k} r_i(W_i-np_{i0}))^2}
{n-\sum_{i \neq k}r_inp_{i0}}.\label{stdd}
\end{equation}
and can be transformed to
\begin{equation}
 X^2_k=\frac{1}{n} \sum_{i \neq k} \frac{r_iW_i^2}{p_{i0}}+\frac{1}{n}
\frac{(n-\sum_{i \neq k}r_iW_i)^2}{1-\sum_{i \neq k}r_i
p_{i0}}-n\label{stdd2}
\end{equation}
that is convenient for numerical calculations. Asymptotically the
vector $\textbf{W}$ has a normal distribution $\mathcal{N}(n
\textbf{p}_0,\Gamma_k^{1/2})$   \cite{robins} and therefore the test
statistic (\ref{stdd}) has $\chi^2_{m-1}$ distribution if hypothesis
${H_0}$ is true. Notice that for usual histograms when $r_i=1$,
$i=1,\ldots, m$ the statistic (\ref{stdd}) is Pearson's chi-square
statistic. The expectation value of statistic (\ref{stdd}) is equal
to
\begin{equation}
\textrm E \,X^2_k=\frac{1}{n} \sum_{i \neq k} \frac{r_i \textrm E
\,W_i^2}{p_{i0}}+\frac{1}{n} \frac{n^2-2n\sum_{i \neq k}r_i \textrm
E \,W_i+\textrm E\,(\sum_{i \neq k}r_iW_i)^2}{1-\sum_{i \neq k}r_i
p_{i0}}-n.
\end{equation}
According  (\ref{ew}) and (\ref{ga11})
\begin{equation}
\textrm E \,W_i^2=np_{i0}/r_i-np_{i0}^2+n^2p_{i0}^2
\end{equation}
and
\begin{equation}
\textrm E\,(\sum_{i \neq k}r_iW_i)^2=-n(\sum_{i \neq k}r_i
p_{i0})^2+n^2(\sum_{i \neq k}r_ip_{i0})^2+n\sum_{i \neq k}r_ip_{i0}
\end{equation}
then
\begin{equation}
\begin{split}
\textrm E \,X^2_k=m-1+(n-1)\sum_{i \neq k}r_ip_{i0}-n\quad\quad\quad\quad\quad\quad
\quad\quad\quad\quad\quad\quad\quad\quad\quad\quad\\
+\frac{n-2n\sum_{i \neq k}r_ip_{i0}-(\sum_{i \neq k}r_i
p_{i0})^2+n(\sum_{i \neq k}r_ip_{i0})^2+\sum_{i \neq
k}r_ip_{i0}}{1-\sum_{i \neq k}r_i p_{i0}}\quad\\
=m-1\quad\quad\quad\quad
\quad\quad\quad\quad\quad\quad\quad\quad\quad\quad\quad\quad\quad\quad\quad\quad\quad\quad\quad\quad\quad\:\,
\end{split}
\end{equation}
as for Pearson's test \cite{kendall}.

 Let us now replace $r_i$ with the estimate $\hat
r_i=W_i/W_{2i}$ and denote the estimator of matrix $\Gamma_k$ as
$\hat \Gamma_k$. Then for positive definite matrices $\hat
\Gamma_k$, $k=1,\ldots,m$ the test statistic is given as
\begin{equation}
\hat X^2_k=\frac{1}{n} \sum_{i \neq k} \frac{\hat
r_iW_i^2}{p_{i0}}+\frac{1}{n} \frac{(n-\sum_{i \neq k}\hat
r_iW_i)^2}{1-\sum_{i \neq k}\hat r_i p_{i0}}-n.\label{statil}
\end{equation}

Formula (\ref{statil}) for usual histograms does not depend on the
choice of the excluded  bin, but for weighted histograms there can
be a dependence. A test statistic that is invariant to the choice of
the excluded bin and at the same time is Pearson's chi square
statistics for the usual histograms can be obtained as the median
value of (\ref{statil}) with positive definite matrix $\hat
\Gamma_k$ for a different choice of excluded bin
\begin{equation}
\hat X^2= \textrm {Med }\, \{\hat X_1^2,  \hat X_2^2,  \ldots , \hat
X_m^2\}.\label{stdav}
\end{equation}
Usage of $\hat X^2$ to test the hypothesis $H_0$ with a given
significance level is equivalent to making a decision by voting.

Use of the chi-square tests is inappropriate if any expected
frequency is below 1 or if the expected frequency is less than 5 in
more than 20\% of bins \cite{moore}. This restriction known for
usual chi-square test is quite reasonable for weighted histograms
also and helps to avoid cases when matrix $\hat \Gamma_k$ is not
positive definite.

Notice also that for case $W_i=0$  the ratio $\hat r_i$ is
undefined. The average value of this quantity for nearest neighbors
bins with non-zero  bin content can be used for approximation of
undefined  $\hat r_i$.

\section{The test for histograms with unknown normalization}

In practice one is often faced the case that a histogram is defined
up to an unknown normalization constant $C$. Let us denote a bin
content of histograms without normalization as $\check{W_i}$, then
$W_i=\check{W_i}C$, and the test statistic (\ref{stdd2}) can be
written as
\begin{equation}
 X^2_k=\frac{C}{n} \sum_{i \neq k} \frac{\check{r}_i\check{W}_i^2}{p_{i0}}+\frac{1}{n}
\frac{(n-\sum_{i \neq k}\check{r}_i\check{W}_i)^2}{1-C^{-1}\sum_{i
\neq k}\check{r}_i p_{i0}}-n,\label{stddc}
\end{equation}
with $\check{r_i}=Cr_i$. An estimator for the constant $C$ can be
found by minimization of (\ref{stddc}).
 The normal equation for (\ref{stddc}) has the form
\begin{equation}
 \sum_{i \neq k} \frac{\check{r}_i\check{W}_i^2}{p_{i0}}-
   \frac{(n-\sum_{i \neq k}\check{r}_i\check{W}_i)^2 }
   {(C-\sum_{i \neq k}\check{r}_i p_{i0})^2} \sum_{i \neq k}\check{r}_i p_{i0}=0\label{normal}
\end{equation}
with two solutions
\begin{equation}
\hat C_k=\sum_{i \neq k}\check{r}_ip_{i0}\pm \sqrt{\frac{\sum_{i
\neq k}\check{r}_ip_{i0}}{\sum_{i \neq
k}\check{r}_i\check{W}_i^2/p_{i0}}}(n-\sum_{i \neq
k}\check{r}_i\check{W}_i), \label{const}
\end{equation}
where $\hat C_k$ is an estimator of $C$. We choose the solution with
the positive sign  because it converges to a constant $C=1$ for the
case of a usual histogram, while the  solution with negative sign
does not. Substituting (\ref{const}) to the (\ref{stddc}) we get the
test statistic
\begin{equation}
{\check{X}}^2_k=\frac{\hat C_k}{n} \sum_{i \neq k}
\frac{\check{r}_i\check{W}_i^2}{p_{i0}}+\frac{1}{n} \frac{(n-\sum_{i
\neq k}\check{r}_i\check{W}_i)^2}{1- \hat C_k^{-1}\sum_{i \neq
k}\check{r}_i p_{i0}}-n\label{stddu}
\end{equation}

that   has a $\chi^2_{m-2}$ distribution if hypothesis ${H_0}$ is
valid. Formula  (\ref{stddu}) can be also transformed to
\begin{equation}
\check{X}^2_k =\frac{s^2}{n}+2s, \label{sss}
\end{equation}
where
\begin{equation}
s=\sqrt{\sum_{i \neq k}\check{r}_i p_{i0} \sum_{i \neq k}
\check{r}_i\check{W}_i^2/p_{i0}} - \sum_{i \neq
 k}\check{r}_i\check{W}_i
\end{equation}
that is convenient for calculations.

The final statistic $\hat {\check{X}}^2_k$ is obtained by replacing
$\check{r}_i$ in (\ref{sss}) with the estimate $\hat{\check{
r}}_i=\check{W}_i/\check{W}_{2i}$. As in chapter 2, a test statistic
that is "invariant" to choice of the excluded bin can be obtained as
the median value of (\ref{sss}) for all possible choices of the
excluded bin
\begin{equation}
\hat {\check{X}}^2= \textrm {Med }\, \{\hat {\check{X}}_1^2, \hat
{\check{X}}_2^2, \ldots , \hat {\check{X}}_m^2\}.\label{stdav}
\end{equation}

\section{Evaluation of the tests' sizes and power}

The tests described herein is now  evaluated  with a numerical example.
We take a  distribution
\begin{equation}
p(x)\propto \frac{2}{(x-10)^2+1}+\frac{1}{(x-14)^2+1} \label{weight}
\end{equation}
 defined on the interval $[4,16]$ and representing two so-called Breight-Wigner
  peaks \cite{breit}. Three cases of the probability density function $g(x)$ are
  considered(see Fig.1)

\begin{equation}
g_1(x)=p(x)   \label{prc}
\end{equation}

\begin{equation}
g_2(x)=1/12  \label{flat}
\end{equation}

\begin{equation}
g_3(x)\propto\frac{2}{(x-9)^2+1}+\frac{2}{(x-15)^2+1} \label{real}
\end{equation}
\begin{figure}
\centering \vspace*{-1.9 cm} \hspace*{-0.7 cm}
\includegraphics[width=1.15 \textwidth]{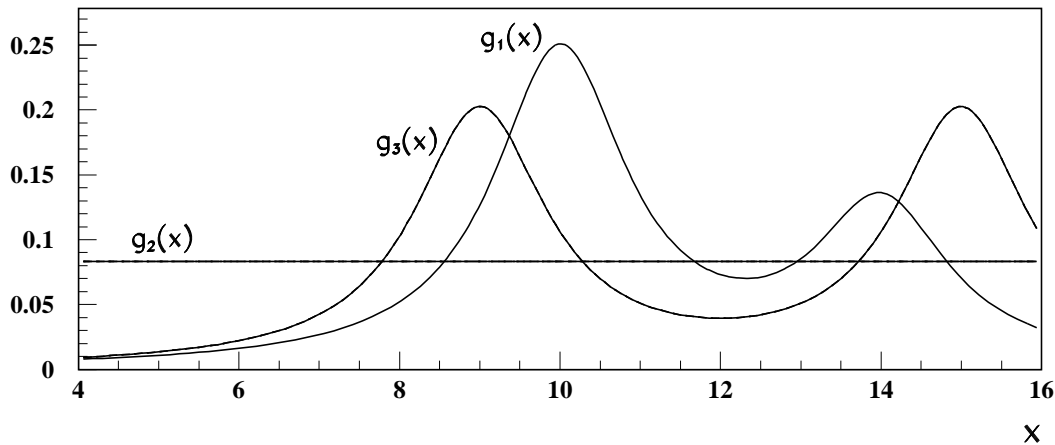}
 \vspace*{-8.5 cm} \caption {Probability density functions $g_1(x)=p(x)$, $g_2(x)$ and
 $g_3(x)$}
 \vspace*{1 cm }
\end{figure}

\begin{figure}
\centering
\vspace*{-1.9 cm}
\hspace*{-1.4 cm}  \includegraphics[width=1.2 \textwidth]{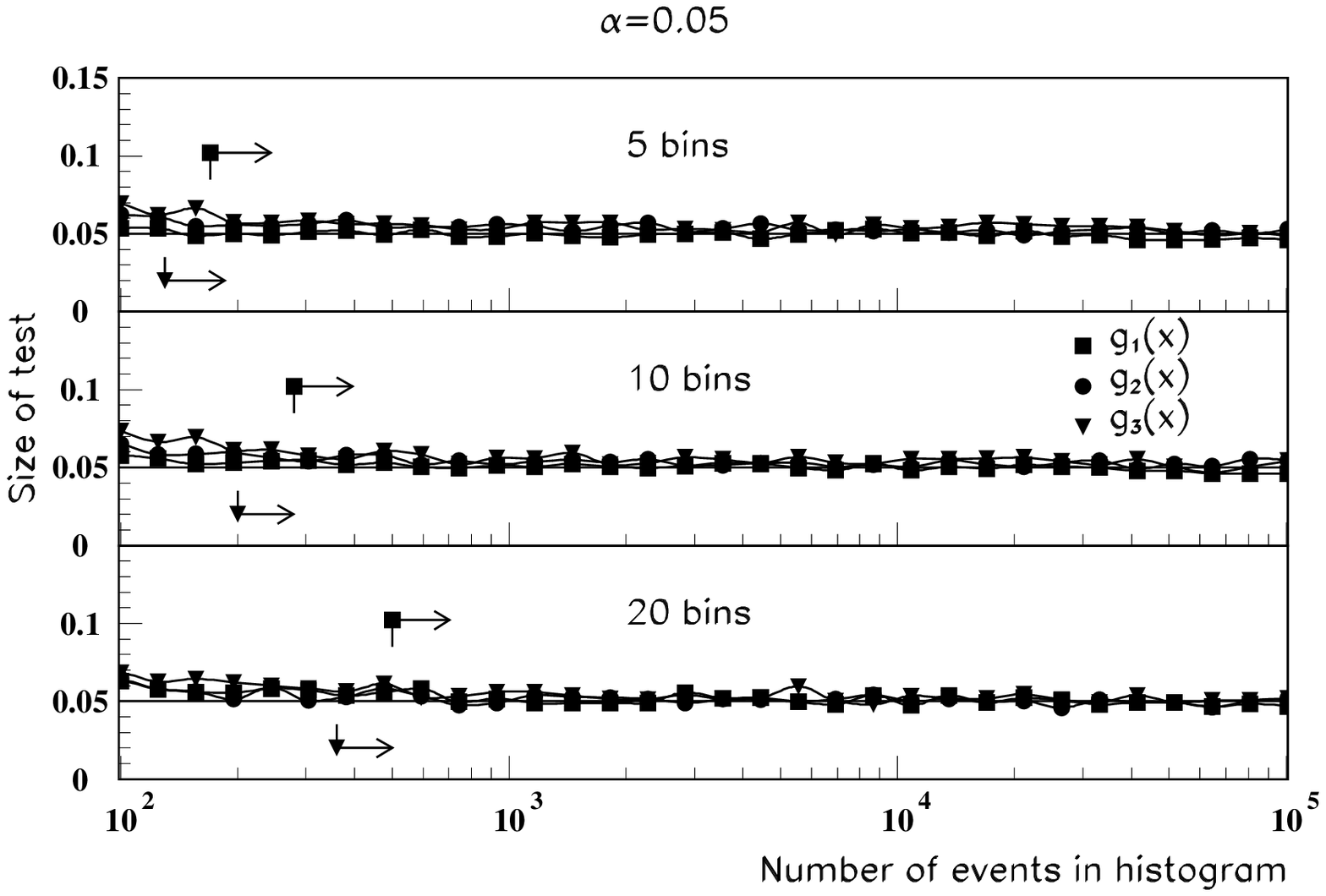}\\
\vspace*{-4.9 cm}
\vspace*{1.9 cm}
\vspace*{-2.9 cm}
\hspace*{-1.4 cm}  \includegraphics[width=1.2 \textwidth]{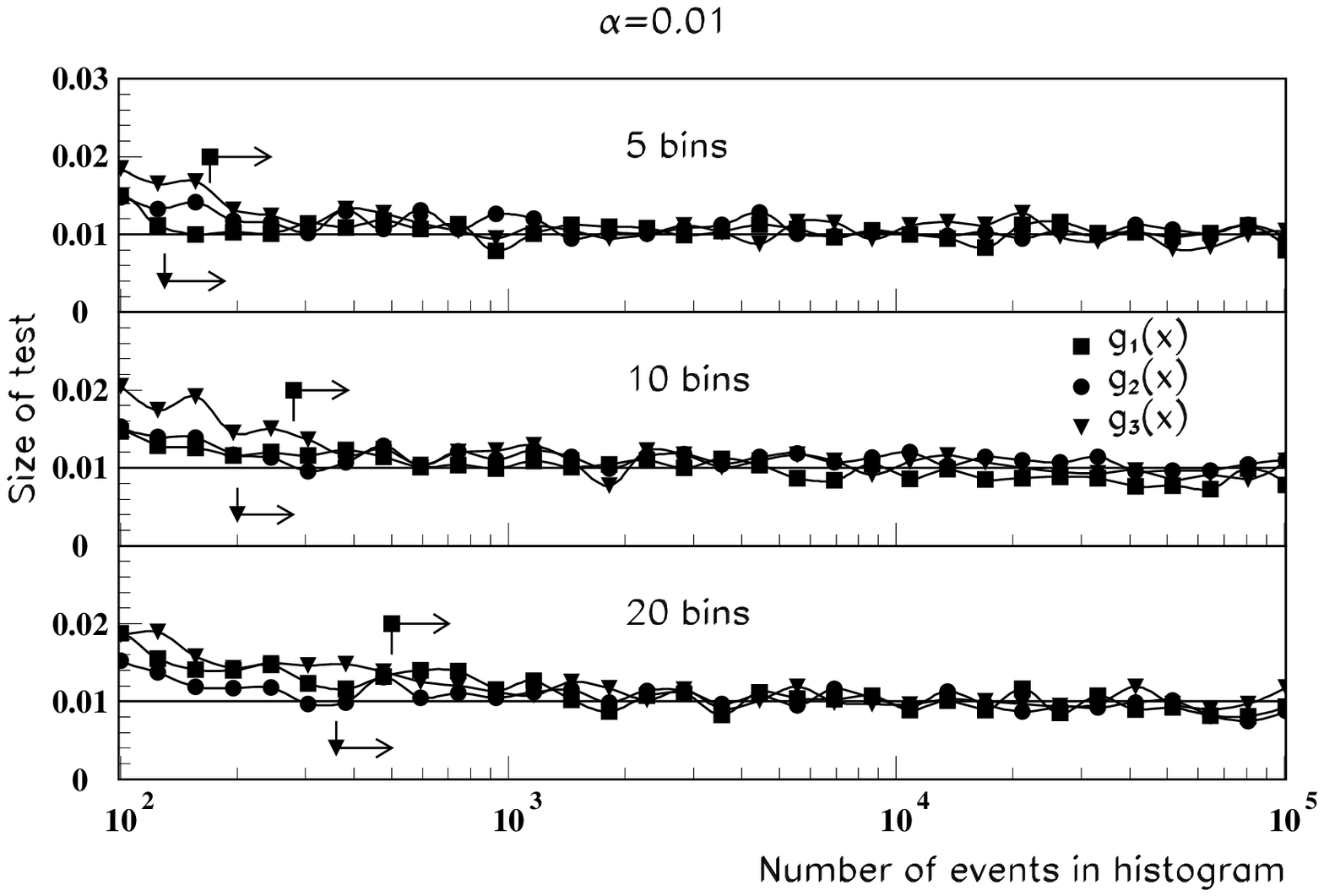}\\
\vspace*{-4.9 cm}
\vspace*{0.5 cm}
\caption {Sizes of the chi-square tests for histograms with different
 weight functions  and different numbers of bins as a function of
 the number of events $n$ in the  histogram. Arrows show regions with
  appropriate number of events in histogram for test application.}
\end{figure}

\begin{figure}
\centering
\vspace*{-1.9 cm}
\hspace*{-1.4 cm}  \includegraphics[width=1.2 \textwidth]{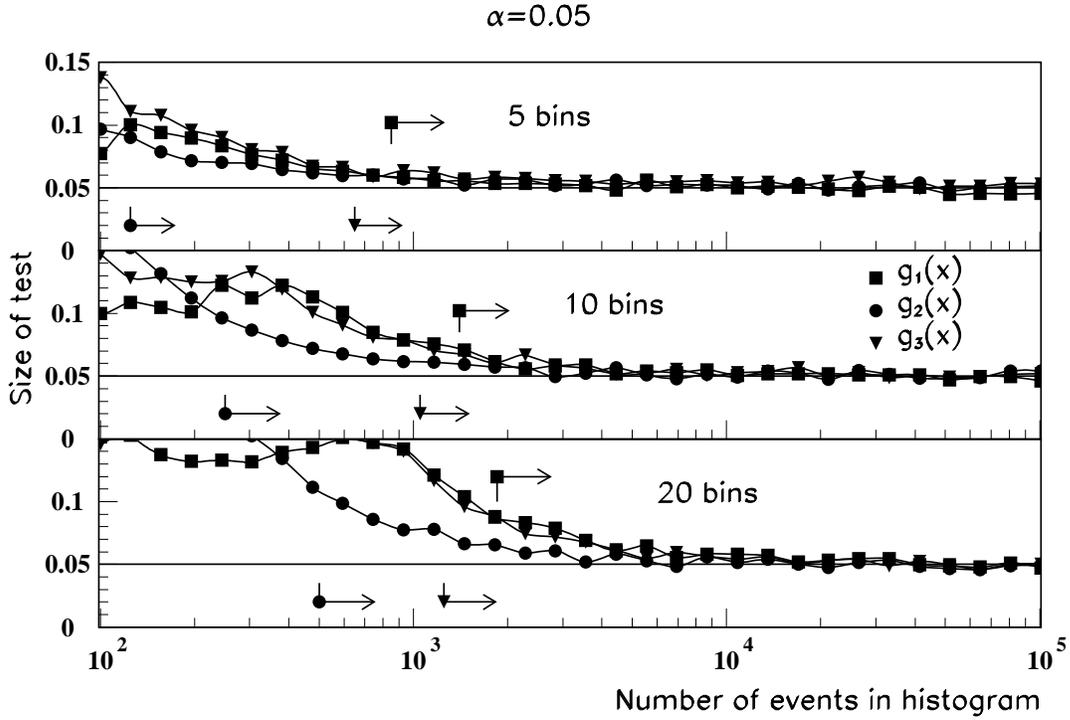}\\
\vspace*{-4.9 cm}
\vspace*{1.9 cm}
\vspace*{-2.9 cm}
\hspace*{-1.4 cm}  \includegraphics[width=1.2 \textwidth]{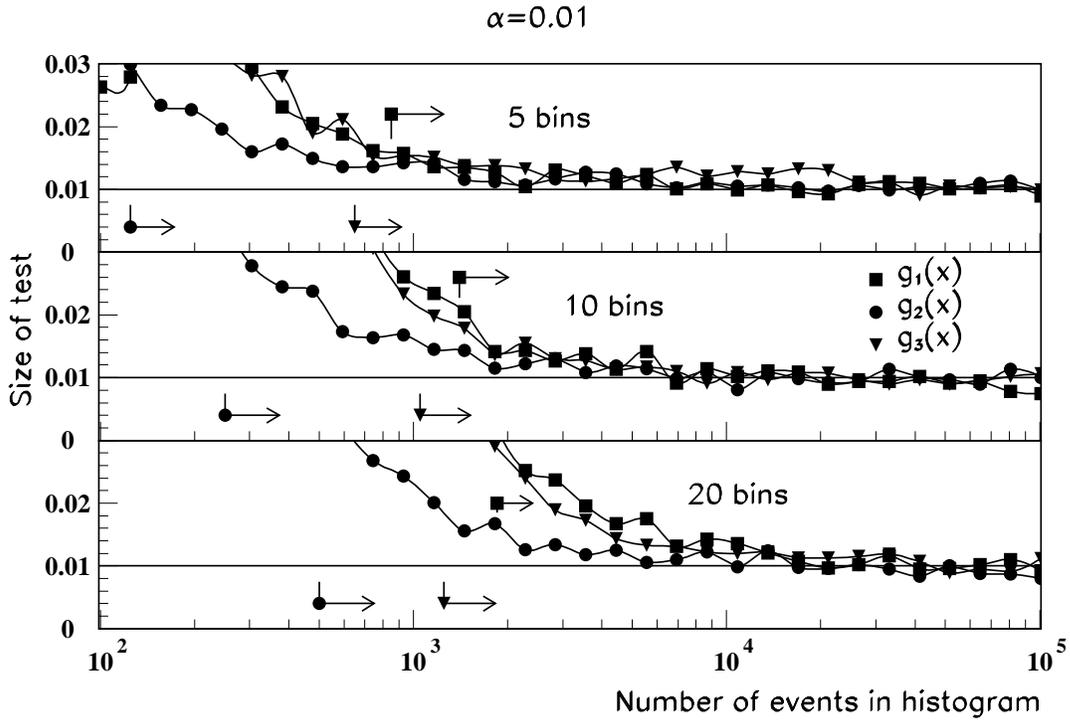}\\
\vspace*{-4.9 cm}
\vspace*{0.5 cm}
\caption {Sizes of the heuristic chi-square test for histograms with
different weight functions  and different numbers of bins as a
 function of the number of events $n$ in the histogram. Arrows show regions
 with minimal number of events in bins of histograms equal to 25.}
\end{figure}

\begin{figure}
\centering
\vspace*{-1.9 cm}
\hspace*{-1.4 cm}  \includegraphics[width=1.2 \textwidth]{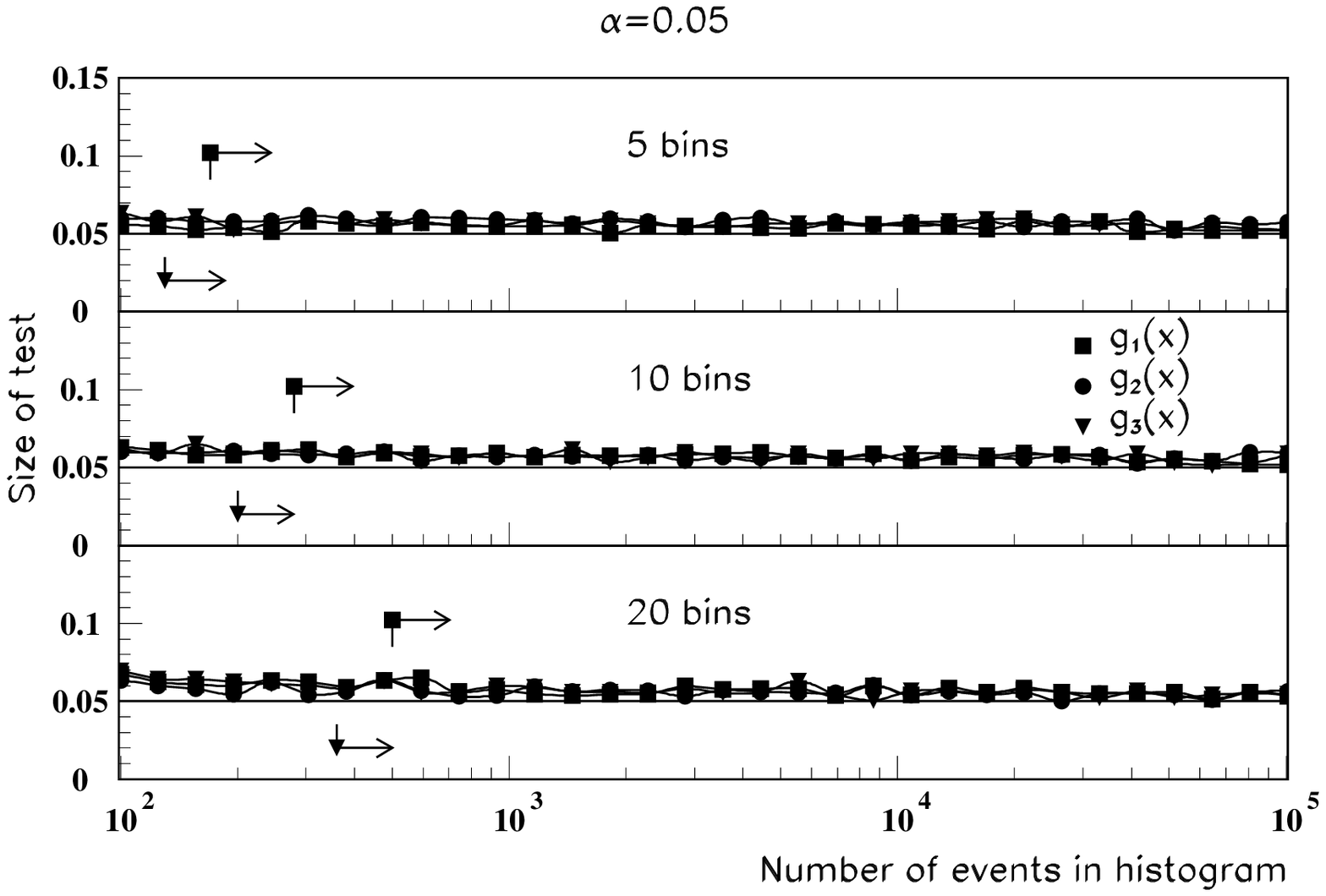}\\
\vspace*{-4.9 cm}
\vspace*{1.9 cm}
\vspace*{-2.9 cm}
\hspace*{-1.2 cm}  \includegraphics[width=1.2 \textwidth]{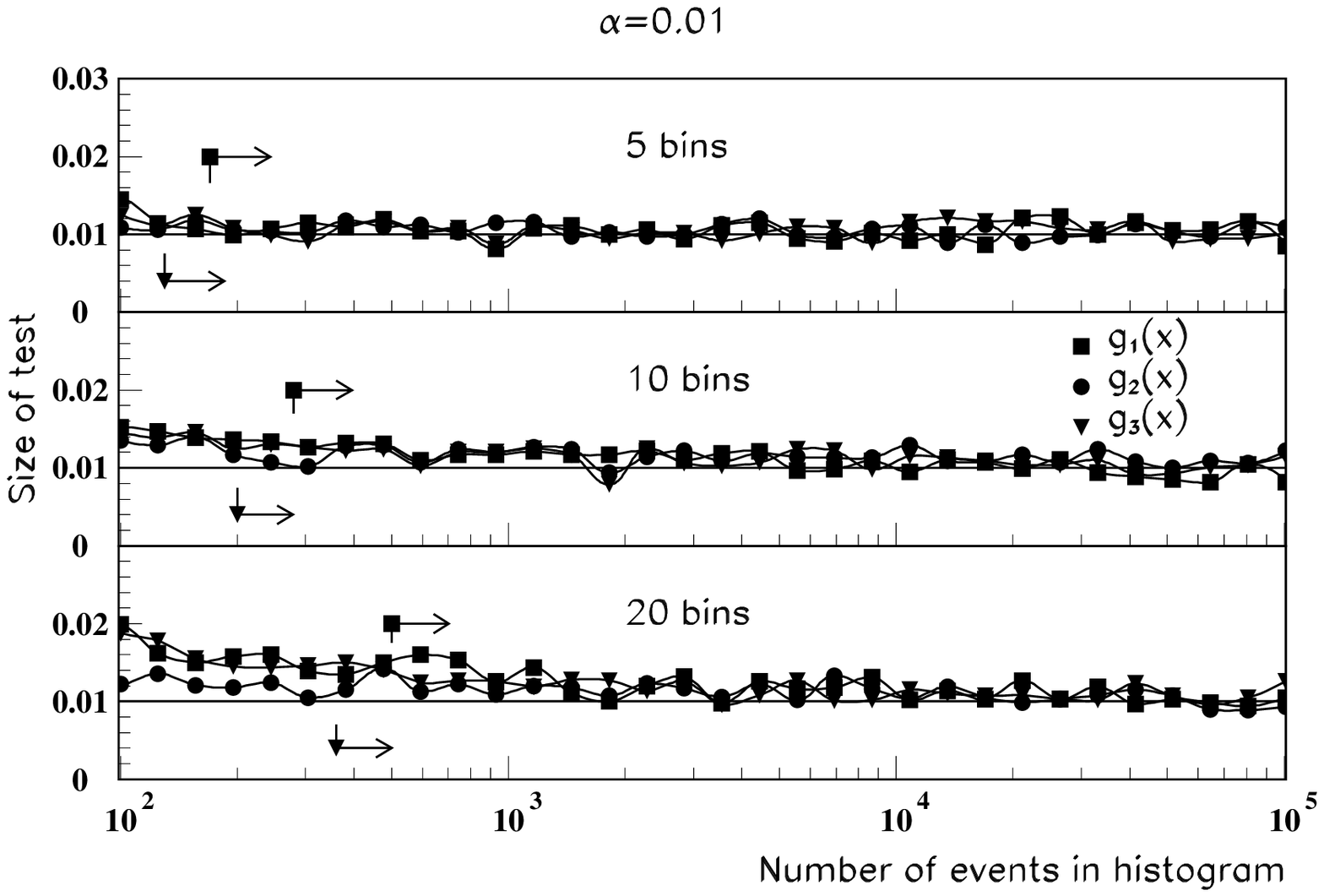}\\
\vspace*{-4.9 cm}
\vspace*{0.5 cm}
\caption {Sizes of tests for histograms with unknown normalization  for different weight functions
 and different numbers of bins as a function of the number of events $n$ in the histogram.}
\end{figure}

\begin{figure}
\centering \vspace*{-1.9 cm}
\hspace*{-0.7 cm}  \includegraphics[width=1.15 \textwidth]{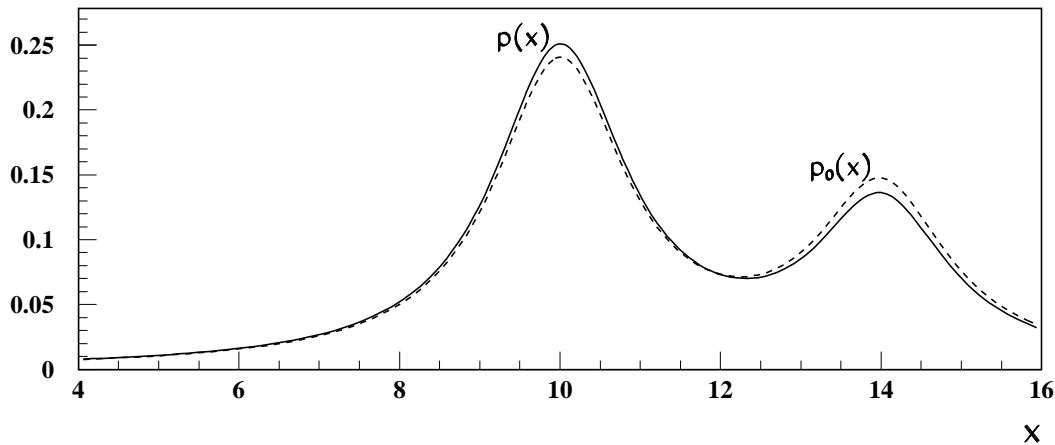}\\
\vspace*{-8. cm} \caption {Probability density function $p(x)$
(solid line)  and $p_0(x)$ (dashed line)} \vspace*{1 cm }
\end{figure}

\begin{figure}
\centering
\vspace*{-1.9 cm}
\hspace*{-1.4 cm}  \includegraphics[width=1.2 \textwidth]{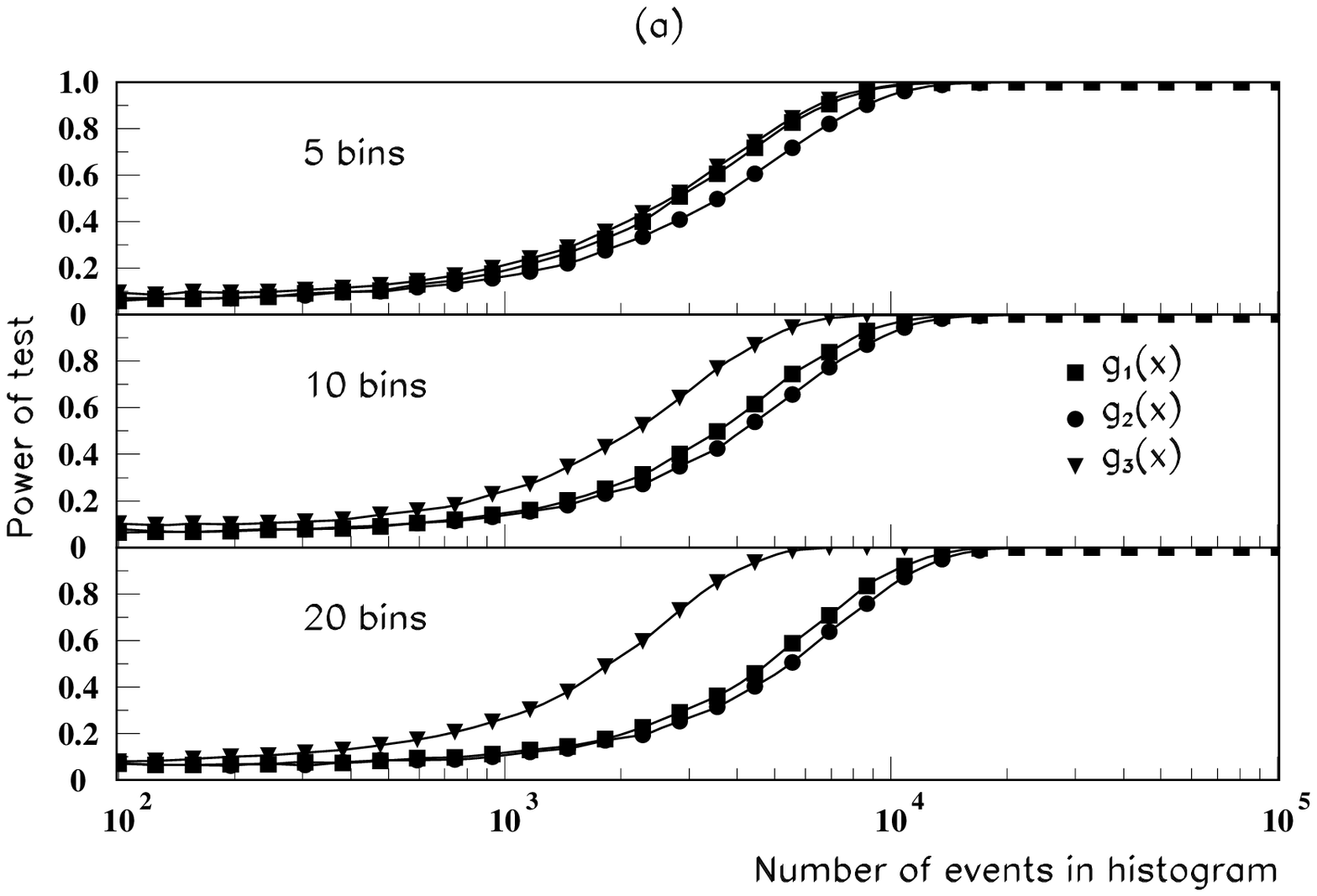}\\
\vspace*{-4.9 cm}
\vspace*{1.9 cm}
\vspace*{-2.9 cm}
\hspace*{-1.4 cm}  \includegraphics[width=1.2 \textwidth]{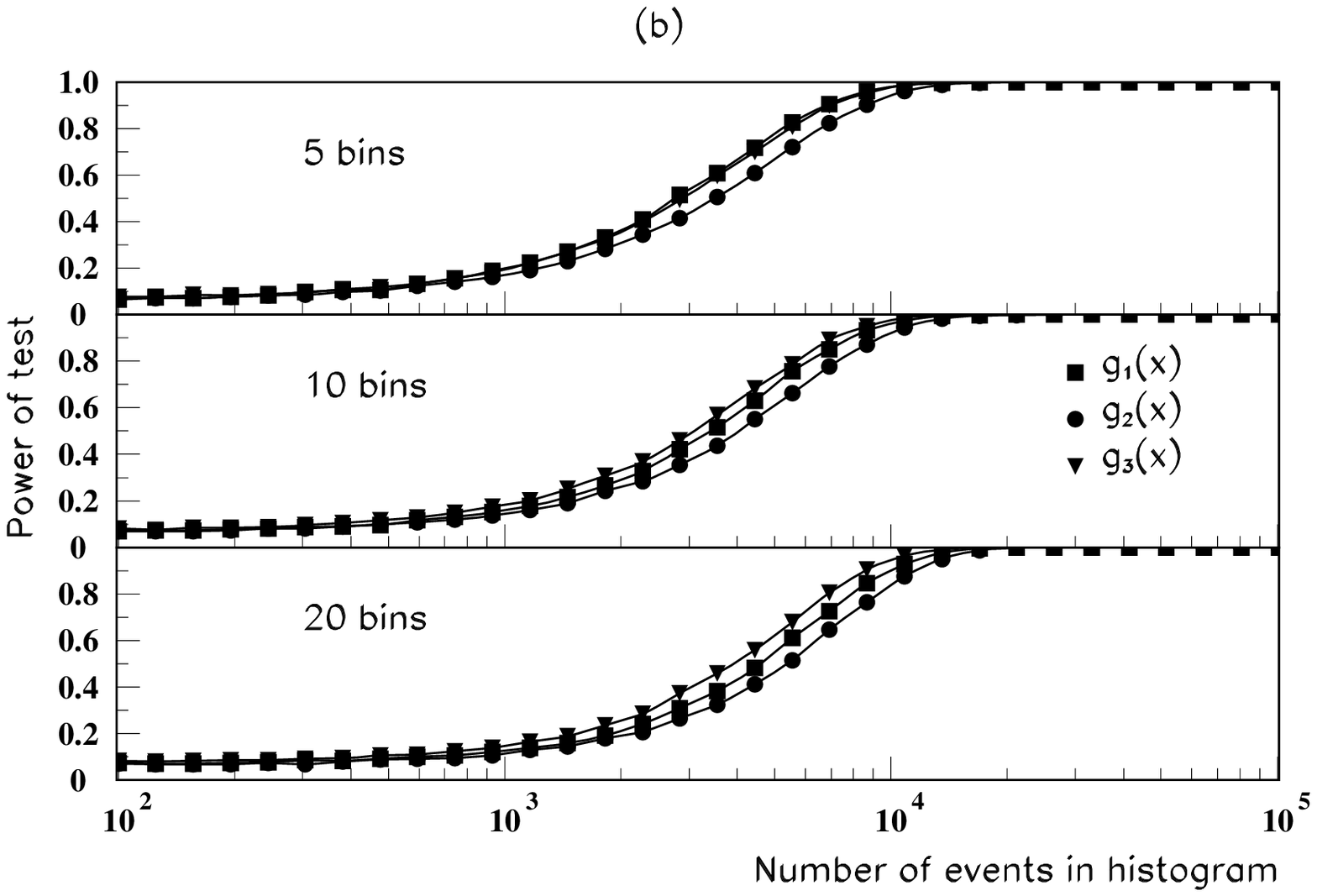}\\
\vspace*{-4.9 cm} \vspace*{0.5 cm} \caption {Powers of chi-square
tests for histograms with different weight functions  and different
numbers of bins as function of the number of events $n$ in the
histogram,  (a) chi-square test generalization, (b) chi-square test
for histograms with unknown normalization; $\alpha =0.05$}
\end{figure}

 Distribution (\ref{prc}) gives an unweighted histogram and the method coincides with  Pearson's chi square test.
 Distribution (\ref{flat}) is a uniform distribution on the interval $[4,16]$.
 Distribution (\ref{real}) has the same form of parametrization as (\ref{weight}), but with
  different values of the parameters.

Sizes of tests for histograms with different numbers of bins  were calculated
for  nominal values of size  equal to $\alpha=0.05$  and for a nominal value
 of size equal to $\alpha=0.01$ (Fig. 2). Calculations of sizes are done using
  the Monte-Carlo method based on 10000 runs. It can be noticed that relative
   deviation of sizes of tests are greater for $\alpha=0.01$ than for
    $\alpha=0.05$. All cases show that test sizes are close to their nominal values
     for large number $n$ of events in the  histogram, and are reasonably
      close to the nominal values for low statistics of events.
The same computation was done for the size of the heuristic test
(see Fig. 3).
  It can be noticed that for large number $n$ of events the sizes of tests
   tend to the nominal value of the test. For small numbers $n$ of events in
   the histograms the sizes of the tests are generally greater then the nominal
     values of tests, although some values of sizes are not shown on the figures
     because they are too big. Comparison of the two tests bring out clearly the
     superiority of the generalization of Pearson's test over the heuristic test.

The same study was done for the chi-square test for histograms  with
unknown normalization. The  results of these calculations are
presented in Fig. 4. Again all cases show that tests sizes are close
to nominal values for large numbers $n$ of events  and reasonably
close to nominal values for low numbers  of events.

The powers of the new chi-square test and the test with unknown
normalization were investigated for slightly different values of the
amplitude of the second  peak of the specified probability
distribution function (see Fig. 5):
\begin{equation}
p_0(x)\propto \frac{2}{(x-10)^2+1}+\frac{1.15}{(x-14)^2+1}. \label{weightr}
\end{equation}

The  results of these calculations are presented in Fig. 6. Notice
that the powers of the tests with unknown normalization are lower
than powers of the normalized test. Comparison of the powers of the
tests for probability density functions $g_2(x)$ and $g_3(x)$ with
powers of the test for function $g_1(x)$ (usual unweighted
histogram) show that the values of  powers are reasonable, as are
the sizes  of the new tests.

\section{Conclusions}

A goodness of fit test for weighted histograms is proposed. The test is a  generalization of Pearson's
chi-square test. Also a goodness of fit test for weighted histograms with unknown normalization is developed.
 Both tests are very important tools in the application of the Monte-Carlo method as well as in simulation studies
  of different phenomena. Evaluation of the sizes and powers of those tests was done numerically  for histograms with
   different numbers of bins, different numbers of events and different weight functions. The same
    investigation was done for the heuristic test used often in practice. Comparison of the results shows the superiority
of the  new tests compared to the  heuristic test.

\ack

 The author is very grateful to Michael Schmelling (Max-Planck-Institut f\"{u}r Kernphysik, Heidelberg)
  for reading the paper  and making constructive comments.

\end{document}